\documentclass[a4paper]{article}

\usepackage{spconf,amsmath,graphicx,cite}

\ninept
\usepackage{amsmath,amssymb,graphicx}
\usepackage[dvipsname]{xcolor}
\usepackage{upgreek,hyperref,enumitem}
\newcommand{\TextTh}{\textsuperscript{th}}

\newcommand{\micron}{\ensuremath{\upmu\text{m}}}
\newcommand{\logo}{{\sf PAINTER}} 		
\newcommand{\ii}{\text{i}} 				
\newcommand{\indic}{\boldsymbol{1}} 			
\newcommand{\Id}{\it{\bf{I}}} 			
\newcommand{\un}{\it{\bf{1}}} 				
\newcommand{\bB}{{\bf{b}}} 			
\newcommand{\ux}{\boldsymbol{x}} 				
\newcommand{\uX}{\boldsymbol{X}} 				
\newcommand{\uy}{\boldsymbol{y}} 				
\newcommand{\uY}{{{\mathrm{\boldsymbol{Y}}}}} 	
\newcommand{\bH}{\boldsymbol{H}} 				
\newcommand{\bh}{\boldsymbol{h}} 				
\newcommand{\uF}{\boldsymbol{F}} 				
\newcommand{\uuF}{{\uF}} 				
\newcommand{\uphi}{\boldsymbol{\varphi}} 		
\newcommand{\upsi}{\boldsymbol{\psi}} 		
\newcommand{\uDphi}{{\boldsymbol\Delta  \boldsymbol{\varphi}}}  		
\newcommand{\bpsi}{\boldsymbol{\psi}} 		
\newcommand{\bdphi}{{\Delta \, \varphi}}  	
\newcommand{\uxi}{\boldsymbol{\xi}} 			
\newcommand{\upow}{\boldsymbol{\gamma}} 	
\newcommand{\datapow}{\boldsymbol{\zeta}} 	
\newcommand{\unoise}{\boldsymbol{\eta}} 	
\newcommand{\uw}{\boldsymbol{\omega}} 	
\newcommand{\TotV}{\text{s}} 			
\newcommand{\SpecS}{\lambda} 			
\newcommand{\tP}{P} 			
\newcommand{\tT}{T} 			
\newcommand{\tS}{S} 			
\newcommand{\tV}{V} 			
\newcommand{\uP}{\boldsymbol{\tP}} 	
\newcommand{\uT}{\boldsymbol{\tT}} 	
\newcommand{\uV}{\boldsymbol{\tV}} 	
\newcommand{\uS}{\boldsymbol{\tS}} 	
 	
\newcommand{\bHtv}{\boldsymbol{H}^\TotV } 		
\newcommand{\mutv}{\mu_s} 		
\newcommand{\bHss}{\boldsymbol{H}^\SpecS} 		
\newcommand{\muss}{\mu_\lambda} 	
\newcommand{\uC}{\boldsymbol{C}} 	
\newcommand{\support}{\boldsymbol{\Pi}} 
\newcommand{\utau}{\boldsymbol{\upsilon}} 	
\newcommand{\uTau}{\boldsymbol{\Upsilon}} 	
\newcommand{\beq}{\begin{eqnarray}}
\newcommand{\eeq}{\end{eqnarray}}
\newcommand{\beqs}{\begin{eqnarray*}}
\newcommand{\eeqs}{\end{eqnarray*}}
\newcommand{\lb}{\left [}
\newcommand{\rb}{\right ]}
\newcommand{\lbr}{\left (}
\newcommand{\rbr}{\right )}

\title{Large scale 3D image reconstruction in optical interferometry}

\name{Antony Schutz$^1$\thanks{The present work was funded by the french ANR project POLCA (ANR-2010-BLAN-0511-02).}
Andr\'e Ferrari$^1$, David Mary$^1$, \'Eric Thi\'ebaut$^2$ and Ferr\'eol Soulez$^2$}
\address{$^1$Lab. Lagrange, Univ.  de Nice Sophia Antipolis, CNRS, Observatoire de la C\^ote d'Azur,  Nice, France\\
$^2$CRAL, Observatoire de Lyon, CNRS, Univ. Lyon 1, \'Ecole Normale Sup\'erieure de Lyon,  Lyon, France.
}

\begin{document}

\maketitle
\begin{abstract} 
Astronomical optical interferometers  (OI) sample the  Fourier transform of the intensity distribution of a source 
 at the observation wavelength.
Because of rapid atmospheric perturbations,
the phases of the complex Fourier samples (visibilities) cannot be directly exploited, and instead linear relationships
between the phases are used (phase closures and differential phases). 
Consequently, specific image reconstruction methods have been devised in the last few decades. 
Modern polychromatic OI instruments are now paving the way to multiwavelength imaging. 
This paper presents the derivation of a spatio-spectral  (``3D'')  image reconstruction algorithm called
\logo \ (Polychromatic opticAl INTErferometric Reconstruction software). 
The algorithm is able to solve large scale problems. It relies on an iterative process, which alternates  estimation of polychromatic images and of  complex visibilities.
The complex visibilities are not only estimated
 from squared  moduli and closure phases, but also from differential phases, which help to better constrain the polychromatic reconstruction.
Simulations on synthetic data illustrate the efficiency of the algorithm.
\end{abstract}
\begin{keywords}
ADMM, irregular sampling, phases estimation, proximal operator, optical interferometry 
\end{keywords}
\section{Introduction}
\label{sec:intro}

The long-standing observation technique called Astronomical Interferometry (AI)
has lead to major discoveries during the last century. In optical wavelengths, the Very Large
Telescope Interferometer in Chile will host in Summer 2015 two next generation 
instruments. In the radio domain, international efforts are being devoted to design and build 
a million-receptor array to be operational in 2020's, the SKA (see \url{https://www.skatelescope.org}). 

In AI, the spatial position of each pair of receivers (telescopes or antennas) defines one of the $N_\bB$ baselines of the telescope/antenna array. 
 In absence of any perturbation and at a given observation wavelength, a pair of receivers with baseline $\bB$ provides a \textit{complex visibility} defined by   $y^\lambda=\widehat{I}(\frac{\bB}{\lambda_n})$, which corresponds to  a  sample of the Fourier spectrum of the intensity distribution of interest  \cite{Thiebaut2010}.


The  sampling function is fully defined by the positions of the interfering receivers and by the observation wavelength.  
Despite the Earth rotation, which changes the geometric configuration of the array w.r.t. the line of sight and thus provides additional samples for observations acquired at different times, 
the sampling of the Fourier space remains  extremely sparse in practice. AI is thus a typical instance
of Compressed Sensing (CS), where the underdetermination of the reconstruction problem (restore an intensity distribution from few
projections in the Fourier space) can be alleviated by exploiting the intrinsic sparsity of astronomical scenes. This fact was early recognized by Astronomers through the CLEAN algorithm \cite{clean}, which is a kind of Matching Pursuit assuming point sources (stars).

In the last ten years, research in the field of interferometric image reconstruction (especially in radio) has been mainly driven by advances in sparse representation models and in optimization methods. Sparse representations using various redundant dictionaries have been used, including synthesis \cite{Li11} and analysis \cite{PURIFY} priors, and combination of both \cite{moresane}.

This paper focuses in Optical Interferometry (OI), for which the problem is more difficult than in radio.
First, OI does not allow accurate measurements of the phases ($\varphi$) of the visibilities because of rapid
atmospheric perturbations. 
This excess of missing information (w.r.t. the radio case, where phases are available) can be  partially recovered by means of linear relationships between the phases, the \textit{phase closures}. This technique combines triplets of phases measured by different telescopes and produces a phase information which is theoretically independent of the atmospheric turbulence.  Second, the number of telescopes in OI arrays is much less than the number of antennas in radio arrays, whence a much sparser Fourier sampling in optical.

In this framework, a classical and well-understood
strategy for image reconstruction is to adopt  an inverse problem approach, where missing information is mitigated, and hopefully compensated for, by \textit{a priori} knowledge \cite{Thiebaut2010}. 
In this case, the image reconstruction algorithm aims at finding an intensity distribution that minimizes a cost function composed of a data fidelity term, which is related to the noise distribution, plus a regularization term and possibly other constraints, which are related to  prior knowledge.
Following this path, various algorithms have blossomed in the last twenty years.
Most of the proposed algorithms rely on gradient descent methods (\textit{WISARD} \cite{Meimon:05}, \textit{BSMEM} \cite{BSMEM}, \textit{MiRA} \cite{MIRA},  \textit{BBM} \cite{BBM}, \textit{IRBis} \cite{IRBis}).
A different approach is used in \textit{MACIM} \cite{MACIM} and in its evolution \textit{SQUEEZE} \cite{baron2012},
which rely on Markov Chain Monte Carlo (MCMC) method.

The case of polychromatic observations, which is under focus here, has recently been made possible through
the advent of multiwavelength interferometers. In this case,
an astrophysical source is described by  an intensity distribution which is a function of the wavelength, and the
inverse problem aims at recovering the spatio--spectral (3D) distribution of the source.
This objective adds to  the intrinsic underdetermination of the OI image reconstruction problem
the computational complexity of solving a large scale inverse problem. 

Polychromatic OI image reconstruction has recently become a very active domain of research.
A spatio-spectral image reconstruction algorithm named \logo \ (for Polychromatic opticAl INTErferometric Reconstruction software) with a  publicly available source code has been proposed in \cite{PAINTER}. 
This algorithm  relies on the visibilities  ($\gamma^2$) as measures of the source power spectrum
and on two types of turbulence independent phases differences:
the phase closures  ($\psi$) at each wavelength
and the differential phases \cite{SelfCal}, which are defined as the  phases relatively to a reference channel. Those constitute additional turbulence-independent observables of the phases  in multiwavelength observation modes.

The objective of this communication is to present a number of improvements brought to the prototype version of
\logo\ algorithm. These improvements  $i)$ regard  the accuracy of the representation model, which involves sparsity in analysis through union of bases and $ii)$ make the algorithm highly scalable. The resulting changes represent an in-depth modification of the original version.
A totally new version of the source code is also publicly available.

The paper is organized as follows: 
section \ref{sec:DataModeling} introduces the notations and data model. 
In section \ref{sec:ModelPhaseDiff} we present  an detailed model of phase relationships. 
Section \ref{sec:InversePb}  tackles the inverse problem approach. 
Section \ref{sec:3Dimages} derives the resulting 3D  reconstruction algorithm.
Performances of the algorithm are presented and analyzed  in section \ref{sec:simus}.

\section{Data model and notations} \label{sec:DataModeling}
Let  $\text{y}^{\lambda_n} $ be the complex visibility  at the spatial frequency  ${\bB}\slash{\lambda_n}$
 , and let  $ \uy^{\lambda_n}$ be the column vector collecting the set of complex visibilities corresponding to all available baselines at wavelength $\lambda_n$. 
The complex visibilities can then be related in matrix form to the parameters by the direct model \cite{Thiebaut2010,admmthiebaut}
\begin{equation}
\uy^{\lambda_n} = \uF^{\lambda_n} ~\ux^{\lambda_n}  \label{eq:yfx}
\end{equation}
where $\uF^{\lambda_n}$ is obtained from a Non Uniform Discrete Fourier Transform (NuDFT) \cite{admmthiebaut} at the spatial frequencies imposed by the geometry of the telescope array and by the observation wavelength $\lambda_n$.
Note that $\uF^{\lambda_n}$ is not an orthogonal matrix.
The previous expression describes the complex visibilities by wavelength. A  compact notation including all wavelengths and baselines is
\begin{align}
\uy &=  \uuF ~\ux , \;  \uuF = \oplus^{N_\lambda}_{n=1} \, \uF^{\lambda_n}  \label{eq:TensorEq} \\
\ux &= \lb {\ux^{\lambda_1}}^\top, \ldots, {\ux^{\lambda_{N_\lambda}}}^\top \rb^\top \nonumber 
\end{align}
where $\uuF $ is a block diagonal matrix with each block referring to the NuDFT at a particular wavelength.
Vector  $\uy$ concatenates the complex visibility vectors ($\uy^{\lambda_n} $ of Eq.~\ref{eq:yfx}) for all wavelengths into a $ N_\bB N_\lambda \times 1 $ visibility vector, with associated moduli  $\upow$ and phases $\uphi$ given by
\begin{equation}
\uy = \lb {\uy^{\lambda_1}}^\top, \ldots, {\uy^{\lambda_{N_\lambda}}}^\top \rb^\top, \; \upow =  | \uy  | , \; \uphi  = \angle\;~\uy \label{eq:phi} 
\end{equation}
In order to analyze the chromatic variation of the visibilities $\uy^{\lambda_n}$ and of the images ${\ux^{\lambda_n}}$ over the $N_\lambda$ wavelengths, 
we also  introduce the  $ N_\bB \times N_\lambda   $ matrix $\uY$ and the $ N^2_x \times N_\lambda $
matrix $\uX$ defined as:
\begin{equation}
\uX = \; \textrm{vec}^{-1} \;\ux , \quad \uY = \; \textrm{vec}^{-1} \;\uy \label{eq:ModMatY} 
\end{equation}
To clarify the use of a matrix notation note that the $n\TextTh$ column of $\uX$, denoted as $\uX_n$, corresponds to the vectorization of the image at the wavelength $\lambda_n$ while the $p\TextTh$ line describes the variation of  pixel $p$ along the wavelengths (\textit{i.e.}, a spectrum).

\section{Model for phase relationships} \label{sec:ModelPhaseDiff}

In the presence of atmospheric turbulence, the beams received at each telescope  are affected by random and different optical paths, which corrupt the phases measurements of the complex visibilities. 
The ``atmospheric corrupted'' visibilities at a given wavelength $\lambda_n$  for the base $\bB_{a,b}$ involving telescopes $a$ and $b$ can be modeled as:
 \begin{equation}
 \text{y}^{\lambda_n}_{a,b} = \gamma^{\lambda_n}_{a,b} \, \exp\lbr\ii\,[  \varphi^{\lambda_n}_{a,b} + \eta^{\lambda_n}_{a} - \eta^{\lambda_n}_{b}   ]\rbr \label{eq:TurbAtmo1}
 \end{equation}
where $\varphi^{\lambda_n}_\cdot$ is the uncorrupted  phases and $\eta^{\lambda_n}_\cdot$ are perturbation terms related to telescopes $a$ and $b$.
To overcome the difficulty of phase estimation, turbulence independent quantities need to be constructed. 

\subsubsection{  Phase closures} \label{sec:T3Phi}
The closure phase allows to get rid of atmospheric effects for triplets of complex visibilities. 
In presence of turbulent measurements, the closure phase ($\psi$) is defined as the phase of the bispectrum \cite{BSMEM}, i.e., the Fourier transform of the triple correlation.
For three baselines $\bB_{a,b}$, $\bB_{b,c}$ and  $\bB_{a,c}$ corresponding to a triplet $(a,b,c)$ of telescopes, the closure phase is defined as:
 \begin{equation}
\psi^{\lambda_n}_{a,b,c}  = \angle\; \text{y}^{\lambda_n} _{a,b }  \text{y}^{\lambda_n} _{b,c }  {\text{y}^{\lambda_n}_{a,c}}^\ast  
= \varphi^{\lambda_n}_{a,b} + \varphi^{\lambda_n}_{b,c}  - \varphi^{\lambda_n}_{a,c} =  \bh^{\lambda_n}_{a,b,c}  \uphi^{\lambda_n}  \label{eq:hphi}
 \end{equation}
 where $\uphi^{\lambda_n}$ is the vector containing all \textit{unperturbed} phases for  wavelength $\lambda_n$, and $\bh^{\lambda_n}_{a,b,c}$ is a sparse row vector with only three non zeros entries that take values $\{1, 1, -1\}$. If $N_t$ denotes the number of telescopes, it is possible to show that
 ${(N_t-1)(N_t-2)}\slash{2}$  independent closure phases per wavelength  are
 available\cite{AdvancedImaging}.
 
 \subsubsection{Differential phases} \label{sec:DiffPhi}
  For  one baseline $\bB_{a,b}$, differential phases ($\Delta \varphi$) measure the phase evolution in wavelength   with respect to a reference phase channel.
 Because the phase turbulence terms  on each telescope  $\eta^{\lambda_k}_{\cdot}$ and $\eta^{\lambda_\text{ref}}_{\cdot}$ are, to a first approximation, independent of the wavelength \cite{DiffVis2012} and
the differential phases  defined by
 \begin{equation}
 \bdphi^{\lambda_k, \lambda_\text{ref}}_{{a,b}}  = \angle\; \text{y}^{\lambda_k} _{a,b }-\angle\; \text{y}^{\lambda_{\textrm{ref}}} _{a,b } =  \varphi^{\lambda_k}_{a,b}  - \varphi^{\lambda_\text{ref}}_{a,b} =  \bh^{\lambda_k, \lambda_\text{ref}}_{a,b}  \uphi  \label{eq:phaseDiff}
 \end{equation}
 are essentially not  affected by the atmospheric perturbation.
The reference channel can be chosen as one of the available channels. 
In this case, $N_\lambda -1$ independent differential phases are available per baseline.
Without loss of generality, we chose $\lambda_1$ as the reference channel and  $\bdphi^{\lambda_k, \lambda_1}_{{a,b}}  =  \bh^{\lambda_k, \lambda_1}_{a,b}  \uphi$, 
where $\bh^{\lambda_k, \lambda_1}_{a,b}$ is a sparse row vector with only two non zeros entries that take values $\{1, -1\}$.

 \subsubsection{Model for all phase relationships} \label{sec:PhaseModelf}
The combination of the  differential phases and phase closures  into a global model will improve the phase estimation:
 indeed, the phase closures constrain the phases of a triplet of bases at a fixed wavelength, while the differential phases constrain the phases  dependence in wavelength for a given base. To derive this model, we denote by
$$\bH^{\uDphi}= \lbr - \un_{(N_\lambda -1)} \otimes \left.  \Id_{N_\bB}  \right|   \Id_{(N_\lambda -1)\times N_\bB} \rbr$$ the matrix  concatenating all vectors $  \bh^{\lambda_k, \lambda_1}_{a,b}$ of Eq.\,\ref{eq:phaseDiff} in its rows. Similarly, 
  $\bH^{\bpsi} = \Id_{N_\lambda} \otimes \bH^{\lambda_\textrm{1}}$ is a block diagonal matrix that replicates the matrix $ \bH^{\lambda_1}$  concatenating the vectors
$\bh^{\lambda_n}_{a,b,c}$ of Eq.\,\ref{eq:hphi} in its rows. The information from the phase closures and differential phases can then be collected in a global vector $\uxi$:
\begin{equation}
\uxi = \bH ~\uphi, \quad   \bH=\lb \frac{\bH^{\bpsi}}{\bH^{\uDphi} } \rb, \quad \uxi=\lb \frac{{\bpsi}}{{\uDphi} } \rb \label{eq:phasecomp}
\end{equation}
where $\upsi$ is the vector of all phase closures  and $\uDphi$ the vector of all differential phases.

\section{Inverse problem approach} \label{sec:InversePb}
According to Eq.~\ref{eq:phasecomp} and notations defined in 
Eq. \ref{eq:phi}, a data model for  phases relationships and squared moduli can be written as:
\begin{align}
 \uxi = \bH ~\uphi + \unoise_\xi ,\quad 
 \datapow = \upow^2 + \unoise_\zeta 
\end{align}
where $\unoise_\xi$ and $\unoise_\zeta$ account for noise and modeling errors.
Classical assumptions on their distributions are considered here.
The noise $\unoise_\zeta$ is assumed to be jointly independent and Gaussian \cite{Meimon:05} and
the noise $\unoise_\xi$ is assumed to be jointly independent and marginally  Von Mises distributed \cite{MIRA}.
Writing the opposite logarithm of the joint likelihood of $ \uxi $ and $ \datapow$ leads to
\begin{equation}
g^\text{data}(\ux) = \alpha \,  g^\zeta(\uy^\gamma) + \beta \, g^\xi(\uy^\phi) \label{eq:cout}
\end{equation}
where $\alpha$ and $\beta$ are relative weighting terms and
\begin{align}
&g^\zeta(\uy^\gamma) = 
\sum_n \frac{1}{\uw_n} \lbr \datapow_n - \upow_n^2 \rbr^2
 \label{eq:Jmod}
, \, \upow = | \uuF \ux | \\
&g^\xi(\uy^\phi) =  - \sum_m \boldsymbol{\kappa}_m \cos{\lbr \bh_m \uphi - \uxi_m \rbr}   \label{eq:Jphase}
, \, \uphi = \angle\;  \uuF \ux 
\end{align} 
Notations $\uy^\gamma$ and $\uy^\phi$ are used above  to underline that the
first  term depends only on the modulus and the second only on the phase of $\uy$, with $\uy = \uF\ux$.
The constant $\uw_n$ is the variances of  $\datapow_n$. The constant  $\boldsymbol{\kappa}_m$ is related to the variance of $\uxi_m$
by $\text{var}(\uxi_m)=1-I_1(\boldsymbol{\kappa}_m)/I_0(\boldsymbol{\kappa}_m)$ where $I_j$ is the modified Bessel function of order $j$.
For a given $\text{var}(\uxi_m)$, $\boldsymbol{\kappa}_m$ is computed solving numerically 
this  equation.

As explained in the introduction,  the problem is severely ill-conditionned owing to the poor coverage of the Fourier space. This requires tackling the image reconstruction as a regularized optimization problem \cite{admmthiebaut}. 
We will adopt here an objective function  of the form: 
\begin{equation}
\ux \leftarrow \underset{\ux \in  \support }{\text{minimize}} \lbr g^\text{data}(\ux) +  f^\text{reg}(\ux) \rbr 
\label{eq:gdatafreg}
\end{equation} 
where the 3D image $\ux$ can be constrained to have a spatially limited support
$\support$. Further constraints such as non negativity can be added in
$f^\text{reg}(\ux)$, which contains all the regularization terms. The support constraint is not included in $f^\text{reg}(\ux)$
for technical reasons related to the ADMM methodology described below.

\subsection{Regularizations and constraints}
OI images are by nature non negative  and sometimes contain sources that are spatially localized.
However, specifying the properties of the object parameters $\ux$ only in terms of non negativity and   spatial support is usually not  a sufficient prior.  
It follows that the use of regularization terms to emphasize some inherent \textit{a priori} knowledge about the image structure is necessary. 
Following the matrix notation  for the 3D object as defined in Eq.~\ref{eq:ModMatY}, 
\logo\  in its current form includes the ridge 
 regularization, motivated by the poor conditioning of 
the NuDFT operator and spatio/spectral regularizations. 
The support constraint is defined by the parameters space $\support$ in Eq.~\ref{eq:gdatafreg}
and the non-negativity constraint by the regularization term $\indic_{\mathbb{R}^+}(\uX)$.
Consequently  the regularization function in Eq.~\ref{eq:gdatafreg} writes:
\begin{equation}
f^\text{reg}(\ux) = \indic_{\mathbb{R}^+}(\uX) \,+\, \frac{\mu_\varepsilon}{2} \| \uX \|_\text{F}^2 	\,+\,    \mutv  \| \bHtv \uX\|_1	  \,+\, \muss \| \bHss \uX^\top\|_1	 
  \label{eq:RegAll}
\end{equation}
$\bHtv$ and $\bHss$ are the matrices associated respectively with the spatial and spectral 
analysis regularizations \cite{admmthiebaut}. 
$\bHtv \uX$ acts on the columns of $\uX$, which are the images at each wavelength processed independently.
$\bHtv$ is a dictionary composed by the concatenation of the first eight orthonormal Daubechies wavelet bases (Db1-Db8) and a Haar wavelet basis. This type of regularization was recently used
in radioastronomy \cite{PURIFY}.
$\bHss \uX^\top$ operates on  the rows of $\uX$ to connect the pixels between wavelengths. In the present work $\bHss$  implements a Discrete Cosine Transform (DCT) but this can be easily replaced by any union of orthogonal bases.
Note that the related matrix $\bHss$ is also an orthogonal matrix.
Finally, $\mu_\varepsilon$, $\mutv$ and $\muss$ are hyper-parameters, which control the weights of the associated regularization terms.
Note that the previous version of \logo\  \cite{PAINTER} was based on total variation regularizations.
Besides the fact that this edge preserving prior does not perfectly match the 
smooth nature of the astrophysical sources, a major drawback of this choice 
comes from the heavy computational cost related to the non-orthogonality  of the
underlying image transform.

\section{3D reconstruction algorithm} \label{sec:3Dimages}
Owing to the unavoidable non convexity of the problem as defined by Eq.~\ref{eq:gdatafreg} (see e.g.  in \cite{Meimon:05}),
the vast majority of image reconstruction algorithms  rely on a descent  optimization principle. So does \logo\, by using the flexibility of the Alternate Direction Methods of Multipliers (ADMM) algorithm, which
was  already used in \cite{admmthiebaut}  to reconstruct  stellar spectra of point 
sources from complex visibilities.  

The optimization problem of Eq.~\ref{eq:gdatafreg} where 
$g^\text{data}(\ux)$ is given by Eqs.~\ref{eq:cout}--\ref{eq:Jphase}
and the regularization term $f^\text{reg}(\ux)$ is given by  Eq.~\ref{eq:RegAll}
is equivalent to:
\begin{align*}
&\underset{\uy^\gamma,\uy^\phi,\uy,\ux,\uP\in\support,\uT,\uS,\uV}{\text{minimize}}  \; 
\alpha \ g^\zeta(\uy^\gamma)+\beta \ g^\xi(\uy^\phi) + \frac{\mu_\varepsilon}{2} \| \uX \|_\text{F}^2 +\\
&  \qquad\qquad\qquad \qquad \qquad \qquad \cdots \indic_{\mathbb{R}^+}(\uP)
+\mutv  \| \uT\|_1+ \muss \| \uV\|_1	 
\\
& \text{s.t.: }  \uy^\gamma =\uy,\,\uy^\phi =\uy,\, \uy= \uuF \ux,\, 
\uT = \bHtv \uX,\, \uV =  \uS \bHss,\, \uS=\uX \nonumber
\end{align*}

Auxiliary variables related to the complex visibilities: $\uy$, $\uy^\gamma$, $\uy^\phi$ have  proper Lagrange multipliers
$\utau_y$, $\utau_\gamma$, $\utau_\phi$ and share the same augmented Lagrangian parameter $\rho_y$.
The auxiliary variables introduced by the regularization, $\uP$, $\uT$, $\uV$, $\uS$, have Lagrange multipliers
$\uTau_P$, $\uTau_T$, $\uTau_V$, $\uTau_S$ and augmented Lagrangian parameters $\rho_P$, $\rho_T$ and $\rho_S$ for $\uV$ and $\uS$.
The $n^{th}$ column of $\uTau_{\cdot}$ is denoted as $\utau_{\cdot}^{\lambda_n}$.
Minimisation of the augmented Lagrangian leads to
solve alternatively and iteratively the following steps:
\begin{enumerate}[label=\Roman*.]
\item \emph{ Minimization w.r.t. $\uy^\gamma$.} Denoting $\tilde{\uy}^\gamma = \uy +\rho_y^{-1} \utau_\gamma$
$$\uy^{\gamma+} = \arg\min_{\uy^\gamma}  \; \alpha \ g^\zeta(\uy^\gamma)  + \frac{1}{2}\rho_y
\|\uy^\gamma - \tilde{\uy}^\gamma\|_2^2$$
This minimization is analytical. It comes down to find the real root of a cubic function using Cardano's method. See \cite{PAINTER}.

\item  \emph{ Minimization w.r.t. $\uy^\phi$.} Denoting $\tilde{\uy}^\phi = \uy +\rho_y^{-1} \utau_\phi$
$$\uy^{\phi+} = \arg\min_{\uy^\phi}  \; \beta \ g^\xi(\uy^\phi)  + \frac{1}{2}\rho_y
\|\uy^\phi - \tilde{\uy}^\phi\|_2^2$$
This minimization is numerically solved using the limited memory BFGS algorithm \cite{lmbfgs}.

\item \emph{ Minimization w.r.t. $\uy$.}
$$\uy^{+} = \frac{1}{3}(\tilde{\uy}^{\gamma+}+\tilde{\uy}^{\phi+}+\uF\ux-\rho_y^{-1}\utau_y)$$
This consensus step leads to complex visibilities reconstruction.

\item \emph{ Minimization w.r.t. $\ux$.} This step operates separately on each wavelength:
\begin{align*}
{\uC^{\lambda_n}}\uX_n^{+} = &
{\uF^{\lambda_n} }^H \lbr \rho_y {\uY}_n^{+}  - 
{\utau}^{\lambda_n} _y \rbr + {\bH^\TotV}^\top  \lbr \rho_\tT \uT_n  - \utau^{\lambda_n} _t  \rbr +  \nonumber \\
& \qquad  \qquad \cdots
  \lbr \rho_\tP\,
{\uP_n} - \utau^{\lambda_n} _p  \rbr   +  
\lbr \rho_\tS \, \uS_n - \utau^{\lambda_n} _s  \rbr   \label{updateX}
\end{align*}
With $\uC^{\lambda_n} = \rho_y {\uF^{\lambda_n} }^H \uF^{\lambda_n} +\eta\,\Id$, where $\eta =  \mu_\varepsilon +  L\rho_\tT+ \rho_\tP + \rho_\tS$.
The Identity matrix in $\uC^{\lambda_n}$ comes from the orthogonality of the 
$L$  wavelet   bases  ($L=9$ here) used  for the spatial regularization.
Computation of the right side term is realised using an inverse NuDFT
and a inverse discrete wavelet transform.
$\uC^{\lambda_n}$ can be inverted using the matrix inversion lemma:
$$
{\uC^{\lambda_n}}^{-1} = \eta^{-1}(\Id - {\uF^{\lambda_n}}^H[\uF^{\lambda_n}{\uF^{\lambda_n}}^H+\rho_y^{-1}\eta\Id]^{-1}\uF^{\lambda_n} )
$$
The number of rows in $\uF^{\lambda_n}$  being small in OI (small number of optical bases),
the inner inverse matrix on the right side of the equality can  be precomputed.
Final computation of $\uX_n^{+}$ is then realized using a NuDFT
and an inverse NuDFT.

\item \emph{ Minimization w.r.t. $\uP$.} Denoting $\widetilde{\uP}_n  = \uX_n^+ + \rho_P^{-1} {\utau_p}^{\lambda_n} $:
$$
\uP^{+} = P_{\mathbb{R}^+}\circ P_\Pi(\widetilde{\uP}) 
$$
where $P_C$ is the projection on the set $C$.

\item \emph{Minimization w.r.t. $\uT$.} Denoting $\widetilde{\uT}_n  = \bH^\TotV\uX_n^+ + \rho_T^{-1} {\utau_t}^{\lambda_n} $:
$$
\uT^{+}_n = \text{Soft}_{\mu_s}(\widetilde{\uT}_n)
$$
where $\text{Soft}_\alpha(\cdot)$ is the soft threshold operator. This step operates separately on each wavelength.

\item \emph{ Minimization w.r.t. $\uS$.}
$$
2\uS^{+} = 
(\uV -\rho_\tS^{-1}\uTau_\tV){\bHss}  +   \uX^+ +\rho_\tS^{-1}\uTau_\tS 
$$
This step operates separately on each voxel. Note that in \cite{PAINTER}
the right term of previous equality was multiplied by the inverse of $\bH^\lambda\bH^{\lambda\top}+\Id$.
The use of an orthogonal DCT reduces this term to $2\Id$.

\item \emph{Minimization w.r.t. $\uV$.} Denoting $\widetilde{\uV}_n  = \uS_n^+ \bH^\lambda + \rho_V^{-1} {\utau_v}^{\lambda_n} $:
$$
\uV^{+} = \text{Soft}_{\mu_\lambda}(\widetilde{\uV})
$$
This step operates separately on each voxel.

\item The Lagrange multipliers are  updated in the standard way, \cite{PAINTER}.
\end{enumerate}
\medskip 
A first acceleration of the proposed algorithm w.r.t. \cite{PAINTER} relies on
the use of orthogonal analysis dictionaries as detailed in the steps IV. and VII.
The bottleneck of \cite{PAINTER} was step IV.
The computational time (matrix--vector multiplication in step IV.) is reduced 
from $\mathcal{O}\lbr N_x^4 \rbr$  to
$\mathcal{O}\lbr 4\,N_x^2 \log(N_x)+ N_\bB(2+N_\bB) \rbr$ in the current version.

Non-equidistant Fast Fourier Transform (NFFT, \url{http://www.nfft.org}) are used to efficiently compute
the NuDFT and inverse NuDFT. Moreover, to take advantage of
the possibility to parallelize the steps IV., VI. and VIII.  w.r.t. the wavelengths or the voxels, the PAINTER algorithm
has been  totally reimplemented in Julia\footnote{\url{http://julialang.org}}. This implementation relies on
OptimPack\footnote{\url{https://github.com/emmt/OptimPack.jl}} for the limited memory 
BFGS algorithm of step II., 
on the NFFT package\footnote{\url{https://github.com/tknopp/NFFT.jl}} and on the
discrete wavelet transform package\footnote{\url{https://github.com/JuliaDSP/Wavelets.jl}}.
Optimisation in step II. is currently the bottleneck of the algorithm.          
The sources are publicly available at  { \url{http://www-n.oca.eu/aferrari/painter/}}.

\section{Simulations} \label{sec:simus}

\begin{figure}[ht!] \vspace{-1mm}
\centerline{\includegraphics[width=.65\columnwidth]{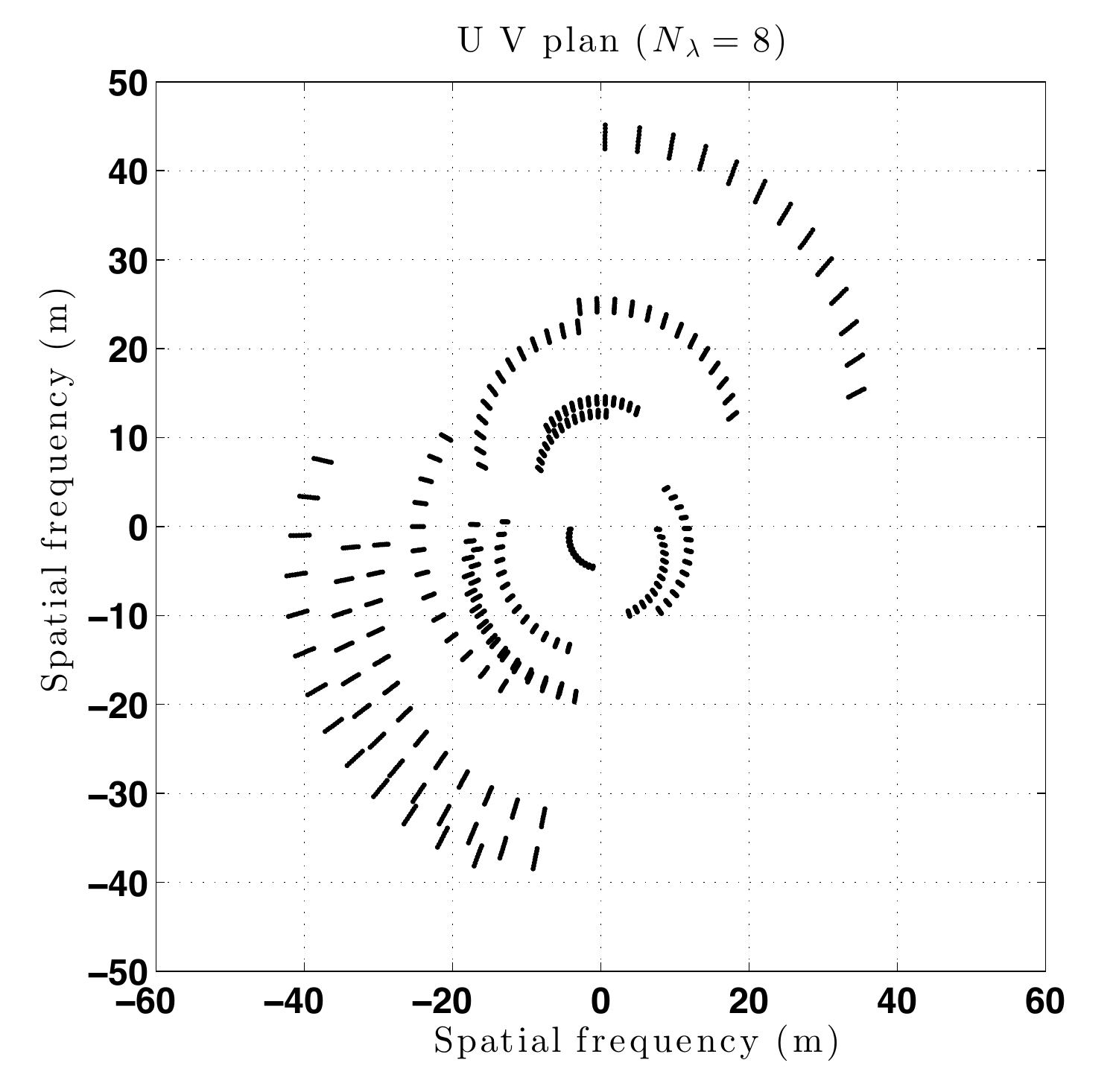}}  \vspace{-3mm}
\caption{Spatial frequencies coverage} \label{fig:uv}   \vspace{-3mm}
\end{figure}

This section presents simulation results obtained with realistic noisy synthetic data. 
The standard deviation (s.d.)  of the noise for the phases  is $\sigma_\varphi$ = 1 rd/SNR (in radians) and the s.d. of the noise for the amplitudes is $\sigma_\gamma$ = $\gamma \slash$SNR. SNR is set to 30 dB in both cases.
The considered instrumental configuration is that of the 2004 International Beauty Contest in Optical Interferometry \cite{IBC04}. The data are acquired in $13$ acquisition epochs and in $8$ equispaced wavelength channels in the range $1.47 \ \micron - 1.56 \ \micron$. The resulting Fourier coverage, including the Earth rotation effect, is shown in Fig. \ref{fig:uv}.

\begin{figure} \vspace{-3mm}
\centerline{\includegraphics[width=.78\columnwidth]{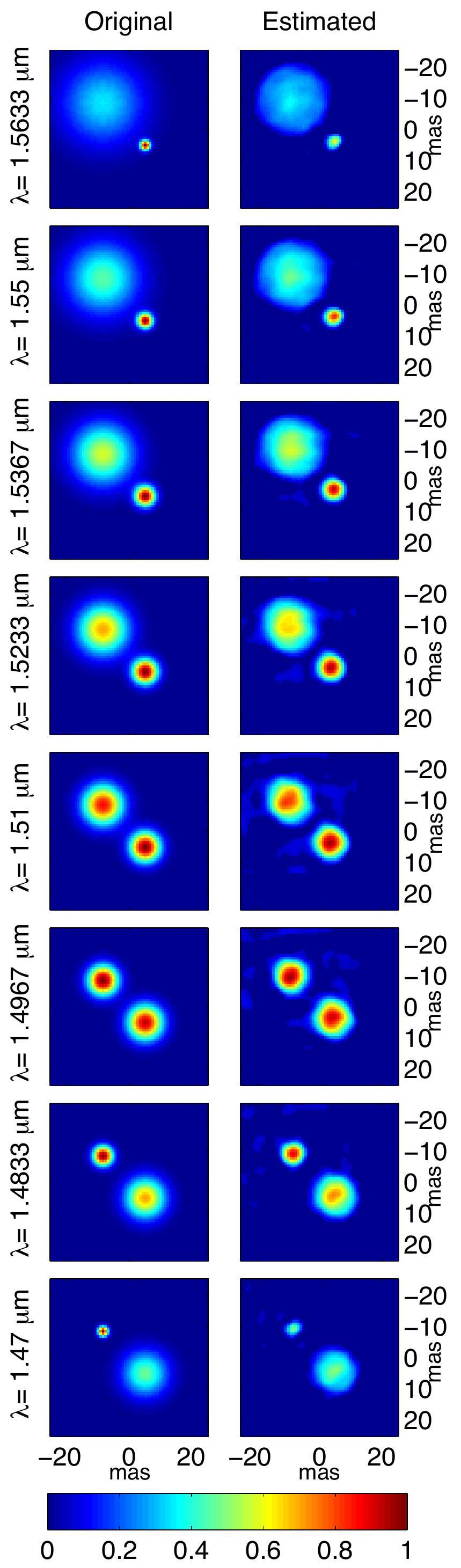}} \vspace{-3mm}
\caption{Original and reconstructed object per channel}
\label{fig:est} \label{fig:slt}
\end{figure}

The considered sources consist in two resolved stars for which the diameter and the brightness distribution 
vary in wavelength (see the left column of Fig.\,\ref{fig:slt}).
 To initialize \logo\, we used  for the $N_\lambda$ images the same image composed of a centered Dirac delta function. 
The size of the reconstructed image at each wavelength is $128\times128$ pixels.
 The algorithm was stopped after 1000 iterations.

\begin{figure}   \vspace{-1mm}
\centerline{\includegraphics[width=.65\columnwidth]{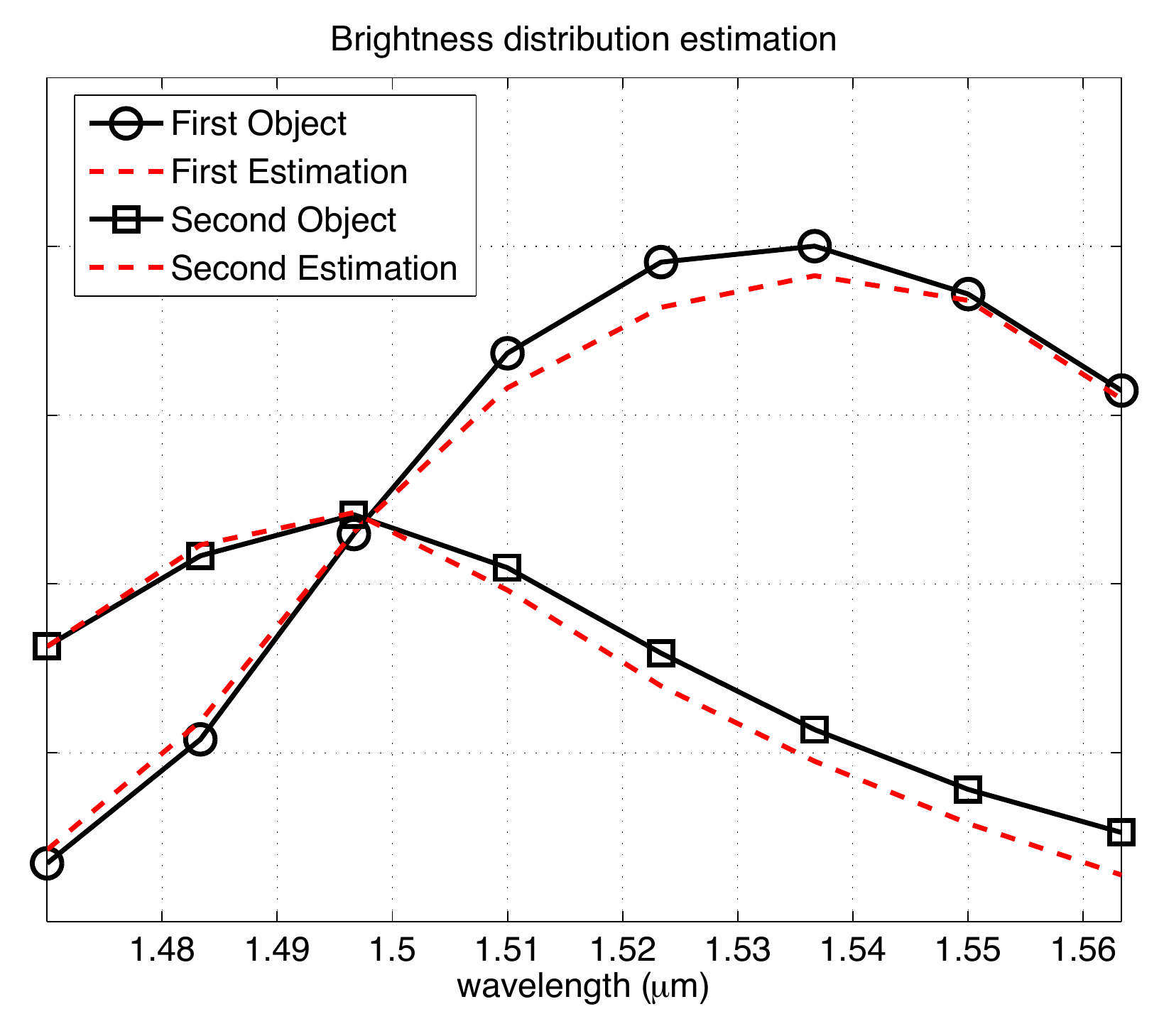}}  \vspace{-3mm}
\caption{Original and estimated brightness distribution} \vspace{-3mm}
\label{fig:dist}  
\end{figure}

The estimated objects are shown in the right column of Fig \ref{fig:est}, which shows that
the shape and the diameter  evolution of the sources are clearly well reconstructed. 
The relative mean square error of the reconstructed image is 7dB with \cite{PAINTER}
and 3.4dB with the proposed algorithm.
The variation of a source's integrated brightness as a function of  wavelength is 
an interesting information \textit{per se} and was thus also investigated (Fig. \ref{fig:dist}). The integrated brightness inside two disks (of diameters  independent of the wavelength and equal to the maximum diameter of each estimated source)  are shown as a function of the wavelength,  both for the  originals and reconstructed sources. Again we find a good match, despite the sparse Fourier coverage.

These results prove the efficiency of the adopted strategy, which
allies redundant dictionaries, algorithmic acceleration and parallel implementation. This
opens the  possibility of providing accurate 3D restoration for large scale problems, involving both high
spatial and spectral resolutions.

\medskip
The authors thank M. Vannier, R. Petrov and F. Millour for fruitful discussion about the use of the differential phase 
and  R. Flamary for the ADMM implementation.


%

\footnotesize
\bibliographystyle{IEEEbib}
\bibliography{refs}

\end{document}